# Integrating Traditional CS Class Activities with Computing for Social Good, Ethics, and Communication and Leadership Skills

**Renato Cortinovis**
Freelance Researcher
Italy
rmcortinovis@gmail.com

**Devender Goyal**
Raytheon Technologies
USA
dg1988@gmail.com

**Luiz Fernando Capretz**
Western University
Canada
lcapretz@uwo.ca

## Abstract
Software and information technologies are becoming increasingly integrated and pervasive in human society and range from automated decision making and social media and entertainment, to running critical social and physical infrastructures like government programs, utilities, and financial institutions. As a result, there is a growing awareness of the need to develop professionals who will harness these technologies in fair and inclusive ways and use them to address global issues like health, water management, poverty, and human rights. In this regard, many academic researchers have expressed the need to complement traditional teaching of CS technical skills with computer and information ethics (computing for social good), as well as communication and leadership skills. In this paper, we describe our goals and some possible class activities we have developed and refined over the past few years with encouraging results, to help CS students understand the potential uses of computing for social good. In these carefully planned project assignments, we seamlessly integrate traditional approaches to develop technical skills with broader professional responsibility and soft skills. We then discuss the lessons learned from these activities and briefly outline future plans.

## 1. Introduction
In the realm of computer science, a new frontier is emerging – one that transcends algorithms and syntax. It's a call from academic researchers, leaders of society, and many public and private institutions like IEEE and ACM, to research and teach computing for social good, ethics, trust, transparency, and building software systems that are appropriate, safe, and reliable, and which can improve the lives of all people (Burton et al., 2018; Charette, 2021; Goldweber et al., 2013; Gotterbarn et al., 2018; Taherdoost et al., 2011). Many academic researchers and industry leaders have also emphasized the need to develop better soft skills in CS students, such as written and verbal communication, collaboration across disciplines, ability to give and take constructive feedback, and empathy, with the aim of better managing conflicts (Capretz, 2014; Capretz et al., 2017; Carter, 2011; Brown et al., 2009; Hazzin & Har-Shai, 2013; Winters & Manshreck, 2020).

We previously described several methods for engaging students and encouraging them to develop these soft skills, highlighting the need to teach computational sustainability, ethics, trust, fairness, and other values in computing (Goyal & Capretz, 2021; Cortinovis, 2021). In this paper, the main goal is to describe class projects and activities dedicated to helping CS students learn concepts of computing for social good, that is, how computer and information technologies can be used to address social issues ranging from health, water resources, poverty, and climate change, to human rights, mainly following the CSG_ED approach described by Goldweber et al. (2013). In order to further improve students' abilities to do good as well as helping to build their professional and personal success, we also integrate some approaches to helping them develop their communication and professional leadership skills.

In the sections below, we first describe Responsible Software Engineering concepts and present a framework for learning computing for social good, computational sustainability and ethical values as important professional skills. We then describe some pilot class activities we carried out, to help students learn these concepts and skills, followed by a discussion of the lessons learned so far. Finally, we hint at the intended follow-up activities.

## 2. Responsible leadership framework
Many researchers are advocating the adoption of social and ethical responsibility by businesses and professionals so that information technologies are used to address social issues in fair, ethical, and inclusive ways. Schieferdecker (2020) has stressed the need to adopt a broader set of values concerns in Software Engineering, which he has entitled 'Responsible Software Engineering,' described as follows:

1. Sustainability by design: In addition to the promotion of privacy, safety and security, and software quality, sustainability concerns like ecological sensitivity for energy and resource efficiency and value sensitivity in data collection and algorithms should be part of software engineering.
2. Techno-social Responsibility: Understanding how digital models could affect the society and shaping the digital business models and solutions according to agreed upon societal principles.
3. Responsible Technology development: Promoting research and development technology that is aligned with the UN sustainable development goals.
4. State-of-the-art Software Engineering: Promoting a sense of societal responsibility in the appropriate use of state-of-the-art software engineering methods and tools that fit the level of software criticality.

Based on our professional experience and academic research, in this paper we are also proposing that a broader set of professional values, responsibilities, and leadership qualities be considered. We propose the following framework and principles for computer science students and professionals to consider:

1. Consider research and development in the areas of computing for social good as well as sustainable computing. That is, harness computational and information technologies to achieve environmental, economic, and societal goals for sustainable development and global good. Many of the global issues that can be addressed by computing professionals relate to health, poverty, climate, legislation, and justice, as well as human rights. Some computational sustainability related areas are sustainable supply-chain systems, disaster management and resilience, energy savings and efficiency, sustainable agriculture, biodiversity and species conservation, and ecosystem informatics (Chatterjee & Rao, 2020).
2. Consider ethics and fairness when designing and developing software and information systems and consider non-functional qualities of software and information systems, such as reliability, safety, and security. Teachers should encourage critical inquiry, reflection, and discussion among participants to help students better prepare for the future (Burton et al., 2018). It is important to help students recognize that multiple stakeholders are impacted by the increasing use of information technologies such as automation of processes, connected systems, social networks, and Artificial Intelligence. In this regard, teachers can also stress the importance of expanding business processes such as risk assessment and risk management which are needed to protect the interests of an organization and its various stakeholders from unethical, and many times unintended, uses of technologies (Clarke, 2019).
3. Be willing to take time to consider wider and long-term impacts of their research and development projects. In this regard, students can be exposed to various frameworks like ethical decision making (Barger, 2008) and RRI (Responsible Research and Innovation) processes of governance (Jirotka et al., 2017), which aim to ensure that the processes as well as the outcomes of the research are aligned with social values by encouraging more inclusive and democratic decision making among various stakeholders.
4. Be professional in their behavior and help each other to learn and grow. As more companies are adopting Agile and Scrum methodologies and becoming aware of servant leadership concepts like empathy, stewardship, healing, and building community (Layton & Ostermiller, 2017), it is imperative that students internalize and practice these concepts. It would be helpful for students to learn 'Virtuous Advocate' values of concern for others and use moral means for social good (Akers et al., 2004).

## 3. Computing for social good

Goldweber et al. (2013), in particular, advocated for the seamless integration of computing educational activities for the social good (CSG_ED) in the traditional CS core activities. Their approach addresses two interrelated issues: the need for students to have skills and attributes that help them contribute to social good outcomes, and motivational issues in students who avoid taking computer science classes as a result of the inaccurate perception that computing careers do not make a difference or contribute to society. Their main objective was indeed to increase the interest in computing by women and other traditionally underrepresented groups, who typically wish to orient their studies in areas where they feel they can contribute to the common good and to the betterment of society in concrete ways. Hence, their approach aims at broadening the view of CS students beyond classical abstract technical domains such as faster algorithms for eigenvalue calculations or sorting numbers, to more socially relevant domains (Schneiderman, 1971).

The key idea is therefore to propose programming assignments that are fully integrated into the core activities typically found in the traditional curriculum but framed in socially relevant ways. They claim that existing research shows that integrating these activities all along the traditional curriculum pathway is more effective than segregating them into separate activities. Additionally, this approach is less intrusive in the sense that it does not force CS educators to reduce the number of technical aspects addressed in the curriculum. Over the years, Goldweber et al. (2019) further refined their model, proposing a taxonomy with four levels of CSG_ED involvement, ranging from redefining an existing problem using a socially-oriented framework, to tackling real world problems with tangible real-world benefits. The main difficulty of this approach seems to be finding a way to avoid increasing significantly the complexity of the assignments for the students.

## 4. Toastmasters approach to developing leadership and communication skills

Goyal et al. (2022) proposed modelling some of the class activities on Toastmasters International, which shows promising results in improving the communication and leadership skills of its members. Toastmasters supports its members' leadership and communication skills in the following ways:

1. Participation in club meetings and competitions provides a safe and encouraging environment to help members practice their communication and leadership skills.
2. Every member supports the growth of other Toastmasters members by way of encouragement and mentoring. The biggest reason for the success of Toastmasters is their encouraging and cooperative environment which helps the members overcome their fear of speaking in front of a group.
3. Members take various leadership roles in running the club successfully and contribute to the growth of Toastmasters clubs. The Toastmasters program provides training sessions to members to prepare for these leadership roles.
4. Providing documentation and mentoring to its members for various roles and club activities.
5. Helping members build relationships and cultivate a sense of community.

The activities present in Toastmasters meetings help its members to develop their communication and leadership skills by giving several opportunities to practice these skills, as follows:

1. Prepared speeches: Each member is provided opportunities to prepare and present speeches in front of other members in a cooperative and encouraging environment. As members gradually become immersed in the encouraging and dynamic environment and listen and learn from other people making speeches, they overcome their fear of speaking in front of a group. Giving a prepared speech also helps a member improve his/her planning, organization, and time management skills.
2. Evaluation and feedback: Giving and receiving feedback is a very important skill. It requires excellent listening skills as well as empathy. Toastmasters training material provides helpful guidelines to its members for providing feedback by first pointing out what the speaker did well and then how he/she can be challenged to improve. The speaker is evaluated on the structure and contents of the speech, as well as delivery skills like the use of effective gestures, body language, vocal variety, and eye contact.

3. Impromptu speaking: Impromptu speaking is called "Table Topics" in Toastmasters. One member prepares the questions and then randomly selects other members to answer the questions by ideally speaking between 1 to 2.5 minutes. Besides improving public speaking skills, it helps members to know more about one another, build relationships, and have some fun time.

## 5. Activities undertaken

We started about three years ago by experimenting with the Toastmasters approach with specific pathways on code of conduct and sustainability (Goyal et al., 2022). Students were instructed to read relevant documentation about these subjects, to prepare a personalized presentation of the main issues, and to discuss it in a Toastmasters like setting. The Toastmasters approach worked well, in the sense that the students engaged actively in the activity, but a preliminary evaluation showed that many students perceived these activities as not sufficiently relevant, and that they preferred to focus on more traditional technical aspects of computer science. This was likely influenced by the specific target population of a continuing education school for adults specialized in computing. These students were mainly young working adults looking for the quickest possible way to acquire the skills necessary to move into a new career in computer science. A few of them were already certified or had even graduated in different fields, but were looking to acquire more easily employable technical skills. A minority had higher objectives such as deepening their studies in CS at the university level, but again these students were usually quite focused on technical skills.

A possible solution to the negative perception reported by some students comes from the integration of the CSG_ED approach. While one of the main aims and rationales of Goldweber et al. (2013) was to enhance the appeal of computer science for potential students intrigued by social welfare, our aim was, on the contrary, to leverage the already existing strong motivation to study CS, in order to expand their perspectives and engagement with social issues, and acquire a broader understanding and appreciation of ethics and values in CS. Based on this strategy, we refocused the activities by seamlessly integrating CSG_ED project assignments into the computer programming activities planned in the traditional curriculum, followed by a scaled down version of the Toastmasters approach to share and discuss the outcomes. The project assignments proposed were as follows:

1. An adapted version of a RoboCat chasing radioactive mice: a simulation of a robot seeking out and catching contaminated mice in a nuclear energy facility following radioactive leakages (Goldweber et al., 2013). The students of an introductory first year Computer Science class first developed a solution to the basic problem of a cat with a single mouse. In doing so, they applied what they had learned about bi-dimensional arrays, and about functional decomposition, working in Java with simple – though creative in some cases – text-based interfaces (Fig. 1). The most advanced students in the class then considered more complex situations with more stationary mice, and experimented with different strategies such as brute force, greedy techniques, or Brownian movements.

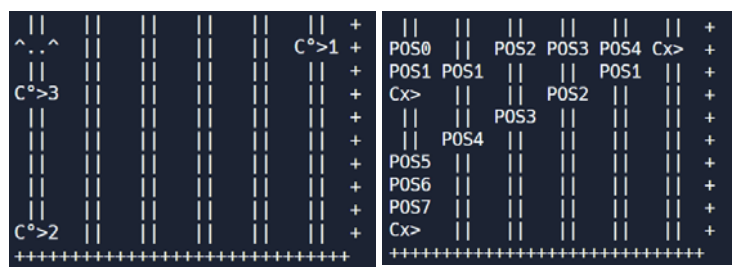

*Figure 1: RoboCat with 3 radioactive mice, greedy strategy (text-based version)*
*Credit: K. D. Faye*

2. As an important spin-off activity, a more advanced student in a final class started developing a graphical interface for the RoboCat world, to be used as a framework so that the students of the first class could concentrate on writing the strategy algorithm but would see the outcomes of their work through a more appealing and motivating interface. This framework will be supported by a few HTML pages clarifying the rationale for the exercise (sustainability, nuclear energy) and some background information. The framework will be published in the open as an OER (Open Educational Resource), with the hope of making it a building block of an OER-enabled open pedagogy activity, where other students will further refine and extend the framework in a self-sustainable cycle – as one of us has already successfully demonstrated in similar contexts (Cortinovis, 2022).

3. Another CSG_ED style project assignment, this time for a more advanced class, involved working on a program to compute the ecological footprint. Students were asked to provide a definition of ecological footprint, document how to compute it, and implement an interactive Java application to actually carry out its computation. Here they applied fundamental object-oriented technical concepts such as inheritance, polymorphism, use of abstract classes and interfaces, advanced static versus dynamic data structures of different complexity – both pre-existing in the library as well as fully developed from scratch, graphical user interfaces with absolute and relative positioning, up to the use of the iterator and composite design patterns (Fig. 2).

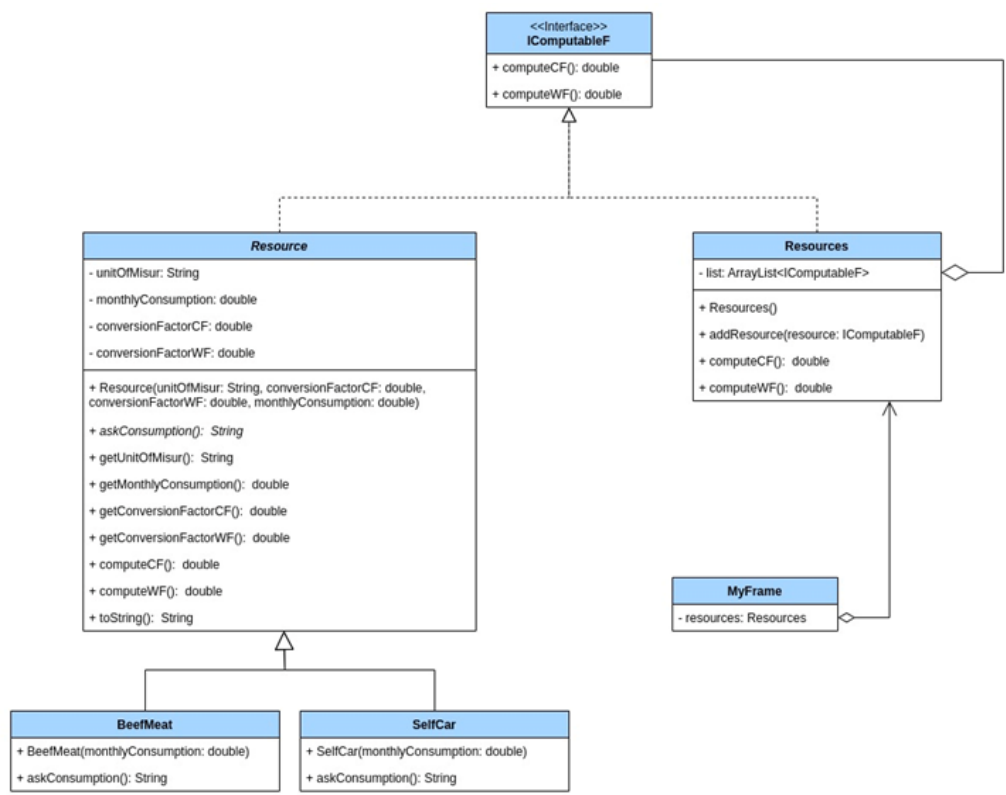

*Figure 2: UML class diagram of the ecological footprint with composite design pattern*
*Credit: N. Invernizzi*

4. Another group of students, studying database design, was asked to analyze the indicators used to track the progress of the Sustainable Development Goals 2030, in order to design a supporting database. Students carried out the analysis and produced an Entity Relationship diagram with the corresponding relational model, implemented it on a relational database management system, and performed a few SQL queries.

Crucially, in order to proactively stimulate critical thinking and concerns about values and social good, all the activities were complemented with a – this time – less preponderant subset of the Toastmasters approach previously piloted in the same classes, mainly including the Table Topics. Some sample questions for Table Topics were:

1. How do you compare the two attempted approaches: reading relevant seed papers, producing a presentation, and discussing it, versus developing a program in a socially relevant domain and briefly discussing it?
2. Do you feel motivated by these assignments in socially relevant contexts, or do you prefer traditional purely technical CS assignments?
3. How does the inclusion of social good concepts affect your perceptions of computer science?
4. How do these activities help you to realize that software can be harnessed to help people in society, improve ecology, and the environment?

In our current pilots, so far, we have done some informational assessment of students' perceptions of the proposed project assignments and teaching approach. Students have definitely found these projects helpful in increasing their motivation and awareness. Yet most students lamented an excess of complexity in the assignments, which – on the contrary – was seen as an interesting challenge by a few more skilled students. The main problem that emerged, in agreement with the literature, is therefore the need to carefully control the complexity of the proposed assignments, which should be personalized according to the students' competencies. The project assignment concerning the computation of the ecological footprint, for example, was carefully refined over a couple of years, to provide plenty of alternative options. These included varying levels of complexity and pedagogical guidance, and optional activities, to improve inclusion by catering to different abilities or motivations of the students. We also found it important to foresee suitable synchronization points, that is, pre-prepared partial baseline solutions, such as a skeleton application or just an UML diagram (Fig. 2), to make sure that nobody got stuck and was unable to progress towards the project goal.

## 6. Discussion – lessons learned

In this section, we summarize the main lessons that we learned from our experience in the last few years, that we feel are worthwhile sharing even at this preliminary stage.

1. We learned in our very first activities that students do not like to address socially oriented concepts or soft skills separately from the activities where they learn technical skills, as they feel they are drifting away from their core objectives – technical orientated for the majority of the students. In contrast, students definitely appreciate the integration of human and social aspects in their standard technical activities, under the form of project assignments with strong technical content complemented with suitable real world and social orientation. This is perfectly in line with the experience and suggestions from the literature (Goldweber et al., 2013).
2. We found that it is necessary to carefully plan the activities with plenty of embedded flexibility so that tackling real world-oriented problems does not excessively increase the difficulty of the proposed assignments. This is again in line with the recommendations from the literature (Goldweber et al., 2013). To help solve this problem in our later activities, we found it convenient to carefully pre-plan different levels of guided support as well as plenty of alternative and optional paths, to allow for educators to quickly react to the potential frustration or disengagement of the students. This proved instrumental to improving the level of inclusion, allowing students with different abilities to feel fully integrated and to fully participate in the activities.
3. We consider it convenient to foresee pre-defined synchronization points, that is, pre-prepared intermediate/partial solutions to periodically guarantee that everyone is on the same page and can progress towards the project goal. These synchronization points need to be frequent enough, so that they can be used to timely bring back on track those individual students who are drifting too far away from the intended path. A timely and personalized intervention in these cases avoids the risk that those students feel frustrated, if they are left drifting away for too long, having to set aside their work later on.
4. To keep the proposed activities at the desired level of complexity, even when tackling real world-oriented problems which may require technical skills beyond the level possessed by the students, we found it useful to develop suitable supporting frameworks, for example implementing complex graphic visualizations. This helps to simplify the proposed assignments

     while retaining their realism, and therefore to maximize students' satisfaction and understanding of the whole picture.

5. In our experience the best approach to developing these frameworks is to propose them to students of higher-level classes as non-disposable assignments, motivating them to contribute to the common good by developing software as OER, that will facilitate and improve the learning and understanding of their peers. As reported by Cortinovis (2021), students deeply appreciate the possibility to personally contribute to the common good through this Open Pedagogy approach: engaging in these activities, indeed, is worthwhile even if just considering their exceptional motivational impact. Moreover, this method ensures sustainability by allowing the resulting OER, originating from what may seem like one-time tasks, to continuously evolve and expand through ongoing Open Pedagogy initiatives.

6. We found that to maximize the impact of these supporting frameworks (OER) and students' motivation, it is important that their effort is shared beyond their local environment, making every possible effort to involve students from other countries and different cultures. This has additional advantages in terms of intercultural understanding and motivation.

## 7. Conclusions

In closing, we advocate for an integrated approach that instills in students the duality of their roles as skilled professionals and conscientious global citizens. We attempted to illuminate a promising avenue for preparing the next generation of computer scientists, armed not only with coding skills but also with the power to wield technology for the betterment of society.

More specifically, we advocated for teaching human and social values and responsibilities as well as soft skills to CS students, and presented some class activities that we attempted and refined over a few years with encouraging results. Suggested class activities include CSG_ED projects related to computing for social good, suitably integrated with Open Pedagogy activities, and further activities modelled after Toastmasters International to help develop student's communication and leadership skills in an encouraging, dynamic, and cooperative environment. In our experience, even short exposures to understanding values and responsibilities and developing soft skills are beneficial and motivating for students, hence we have summarized the main lessons we learned from these activities in the last few years. While we do acknowledge that this is not yet a comprehensive, rigorous research study, we felt it was worthwhile sharing our preliminary results with potentially interested communities of researchers and practitioners.

In the near future, we plan to further study the impact of these activities by further improving, extending, and offering them in collaboration with more faculty members in different contexts, and collecting data for a more rigorous scientific analysis. Additionally, we would like to explore the possibility of establishing a mechanism to crowdsource successful CSG_ED projects as OER, to make it possible to reuse and continuously improve them, ideally in the context of OER-enabled activities (Wiley & Hilton, 2018). If anybody is interested in collaborating, please contact us.

## 8. Acknowledgements
The authors would like to thank S. Camarda for his contribution to the Ecological Footprint application, and all the students who keenly engaged in, and contributed to, the activities described.

## 9. References

Aker, M., Eaton, T.V., & Giacomino, D.E. (2004). Measuring and Changing the Values of Accounting Students. Journal of College Teaching & Learning, 1(4), 63-70. DOI: https://doi.org/10.19030/tlc.v1i4.1937.

Barger, R.N. (2008). Computer ethics: A case-based approach. Cambridge University Press, Cambridge, U.K.

Brown, Q., Lee, F., & Alejandre, S. (2009). Emphasizing Soft Skills and Team Development in an Educational Digital Game Design Course. Proceedings of the 4th International Conference on Foundations of Digital Games, 240-0247. DOI: https://doi.org/10.1145/1536513.1536557.

Burton, E., Goldsmith, J., & Mattei, N. (2018). How to Teach Computer Ethics through Science Fiction. Communications of the ACM, 61(8), 54-64. DOI: https://doi.org/10.1145/3154485.

Capretz, L.F. (2014). Bringing the Human Factors to Software Engineering. IEEE Software, 31(2), 104-106. DOI: https://doi.org/10.1109/MS.2014.30.

Capretz, L.F., Ahmed F., & Silva, F.Q.B. (2017). Soft Sides of Software. Information and Software Technology, 92(2017), 92-94. DOI: https://doi.org/10.1016/j.infsof.2017.07.011.

Carter, L. (2011). Ideas for Adding Soft Skills Education to Service Learning and Capstone Courses for Computer Science Students. Proceedings of the 42nd ACM Technical Symposium on Computer Science Education, 517-522. DOI: https://doi.org/10.1145/1953163.1953312.

Charette, R. (2021). Demanding Fair and Ethically Aligned IT for the Future. IT Professional, 23(3), 46-52. DOI: https://doi.org/10.1109/MITP.2021.3070986.

Chatterjee, D., & Rao, S. (2020). Computational Sustainability: A Socio-Technical Perspective. ACM Computing Surveys, 53(5), 1-29. DOI: https://doi.org/10.1145/3409797.

Clarke, R. (2019). Principles and Business Processes for Responsible AI. Computer Law and Security Review, 35(4), 410-422.

Cortinovis, R. (2021). An educational CPU Visual Simulator. Proceedings of the 32nd Annual Workshop of the Psychology of Programming Interest Group (PPIG).

Cortinovis, R. (2022). Evaluating and improving the Educational CPU Visual Simulator: a sustainable Open Pedagogy approach. Proceedings of the 33rd Annual Workshop of the Psychology of Programming Interest Group (PPIG).

Goldweber, M., Barr, J., Clear, T., Davoli, R., Mann, S., Patitsas, E., & Portnoff, S. (2013). A Framework for Enhancing the Social Good in Computing Education: A Values Approach. ACM Inroads, 4(1), 58-79.

Goldweber, M., Kaczmarczyk, L., & Blumenthal, R. (2019). Computing for the Social Good in Education. ACM Inroads, 10(4), 24-29.

Gotterbarn, D., Brinkman, B., Flick, C., Kirkpatrick, M.S., Miller, K., Vazansky, K., & Wolf, M.J. (2018). ACM Code of Ethics and Professional Conduct. Association of Computing Machinery.

Goyal, D., & Capretz, L.F. (2021). Promoting and Teaching Responsible Leadership in Software Engineering. Proceedings of the 32nd Annual Workshop of the Psychology of Programming Interest Group (PPIG).

Goyal, D., Cortinovis, R., & Capretz, L.F. (2022). A Framework for Class Activities to Cultivate Responsible Leadership in Software Engineering Students. 15th International Conference on Cooperative and Human Aspects of Software Engineering (CHASE 2022), Pittsburgh, PA, USA.

Hazzin O., & Har-Shai, G. (2013). Teaching Computer Science Soft Skills as Soft Concepts. Proceedings of the 44th ACM Technical Symposium on Computer Science Education, 59-64. DOI: https://doi.org/10.1145/2445196.2445219.

Jirotka, M., Grimpe, B., Stahl, B., Eden, G., & Hartswood, M. (2017). Responsible Research and Innovation in the Digital Age. Communications of the ACM, 60(5), 62-68. DOI: https://doi.org/10.1145/3064940.

Layton, M.C., & Ostermiller S.J. (2017). Agile Project Management for Dummies. John Wiley & Sons, New York, NY, USA.

Schieferdecker, I. (2020). Responsible Software Engineering. Future of Software Quality Assurance, 137-146. DOI: https://doi.org/10.1007/978-3-030-29509-7_11.

Shneiderman, B. (1971). Computer Science Education and Social Relevance. SIGCSE Bulletin, 3(1), 21-24.

Taherdoost, H., Sahibuddin, S., Namayandeh, M., & Jalaliyoon, N. (2011). Propose an Educational Plan for Computer Ethics and Information Security. Procedia-Social and Behavioral Sciences, 28, 815-819. DOI: https://doi.org/10.1016/j.sbspro.2011.11.149.

Wiley, D., & Hilton, J. (2018). Defining OER-enabled Pedagogy. International Review of Research in Open and Distance Learning, 19(4).

Winters, T., & Manshreck, T. (2020). Software Engineering at Google: Lessons Learned from Programming Over Time. O'Reilly Media, Boston, MA, USA.

Yu-Chih, S. (2008). The Toastmasters Approach: An Innovative Way to Teach Public Speaking to EFL Learners in Taiwan? RELC Journal, 39(1), 113-130. DOI: https://doi.org/10.1177/0033688208091143.